# Thermal Transport in Graphene Composites: The Effect of Lateral Dimensions of Graphene Fillers

Sriharsha Sudhindra[1,2], Farnia Rashvand[3], Dylan Wright[1,2], Zahra Barani[1,2], Aleksey D. Drozdov[4], Saba Baraghani[1,5], Claudia Backes[3], Fariborz Kargar[1,2,5,*], and Alexander A. Balandin[1,2]

[1]Phonon Optimized Engineered Materials Center, University of California, Riverside, California 92521 USA

[2]Department of Electrical and Computer Engineering, Bourns College of Engineering, University of California, Riverside, California 92521 USA

[3]Institute of Physical Chemistry, Heidelberg University, Im Neuenheimer Feld 253, Heidelberg 69120 Germany

[4]Department of Materials and Production, Aalborg University, Fibigerstraede 16, Aalborg 9220 Denmark

[5]Department of Chemical and Environmental Engineering, University of California, Riverside, California 92521 USA

---

[*] Corresponding author: fkargar@ece.ucr.edu; web-site: http://balandingroup.ucr.edu/

## Abstract

We report on the investigation of thermal transport in non-cured silicone composites with graphene fillers of different lateral dimensions. Graphene fillers are comprised of few-layer graphene flakes with lateral sizes in the range from 400 nm to 1200 nm and number of atomic planes from one to ~100. The distribution of the lateral dimensions and thicknesses of graphene fillers has been determined *via* atomic force microscopy statistics. It was found that in the examined range of the lateral dimensions the thermal conductivity of the composites increases with the increasing size of the graphene fillers. The observed difference in thermal properties can be related to the average *gray* phonon mean free path in graphene, which has been estimated to be around ~800 nm at room temperature. The thermal contact resistance of composites with graphene fillers of 1200-nm lateral dimensions was also smaller than that of composites with graphene fillers of 400-nm lateral dimensions. The effects of the filler loading fraction and the filler size on the thermal conductivity of the composites were rationalized within the Kanari model. The obtained results are important for optimization of graphene fillers for applications in thermal interface materials for heat removal from high-power-density electronics.

## I. Introduction

The exfoliation and electrical experiments with graphene [1] motivated studies of other properties of this material, and led to the discovery of unique thermal conductivity of graphene and few-layer graphene (FLG) [2–8]. In recent years, one can witness a transition of graphene and FLG to numerous practical applications, which utilize their thermal properties, including composite coatings, solid heat spreaders and thermal interface materials (TIMs) [9–29]. The use of graphene – FLG mixture as fillers in composites proved to be particularly beneficial. The initial studies found that even with randomly oriented small loading fractions of graphene fillers, the thermal conductivity of epoxy-based composites can be increased by a factor of ×25 [11,30,31]. It was also shown by some of us that by utilizing graphene in non-cured TIMs, the thermal conductivity exceeds that of the best commercial TIMs at relatively low loading fractions of graphene fillers [23]. Reports from different research groups confirmed the potential of graphene fillers for thermal management applications (for comprehensive reviews on the topic see Ref. [32–34]). There is a general consensus that unique heat conduction properties of graphene and FLG are the enablers of such applications [2,8,35–41]. Here and below we adopt a convention that in the thermal context the term *graphene* fillers mean a mixture of single layer graphene and FLG.

Almost all studies of TIMs with graphene fillers are focused on the thermal conductivity and thermal diffusivity values of the resulting composites as the functions of the graphene loading fraction [32–34]. Several studies addressed the issues of the thermal percolation threshold [24,42] and "synergistic" effects [21,43–52]. In the context of various TIMs, it was established that the percolation threshold depends on the filler sizes [53–55]. The existence and the strength of the "synergistic" effect, *i.e.* additional enhancement of the thermal conductivity when single type of filler with dissimilar size distribution or two different filler materials are used, is also known to depend on the size and aspect ratio of the fillers. Despite the importance of the knowledge of how the lateral dimensions of graphene fillers affect the thermal conductivity of the composites, there have been a limited number of studies on this subject [42,56,57]. We are not aware of any reports of the effect of the graphene filler size on the thermal contact resistance of TIMs with the surfaces of interest. Possible reasons for the lack of such data are difficulties in preparation of consistent set of samples with different and verifiable average lateral size in sufficient quantity.

Investigation of thermal transport properties of composites with graphene and FLG fillers that have different lateral dimensions is interesting from both the fundamental science and practical applications points of view. We hypothesize that the lateral dimensions of the graphene fillers should be larger than the *gray* phonon mean free path (MFP) so that the intrinsic thermal properties of graphene fillers are not limited by the phonon – edge scattering. From the other side, the fillers with excessively large lateral dimensions can be technologically impractical owing to the strong bending of the fillers and increased bond line thickness, *BLT*. There are other factors, related to the lateral size of the fillers, which can affect the heat conduction in the composites. They include the Kapitza resistance between the fillers and matrix material [58], the specific surface area, interface area between the filler and matrix [59–63], and the defect density [4,64–66]. The above considerations motivated us to investigate the specifics of heat conduction in TIMs with graphene fillers where the lateral dimensions of the fillers are below, close, or above graphene's gray phonon MFP. A particular emphasis was on selecting a proper sample set with the well-defined average lateral dimensions of the fillers, systematic analysis of the filler size distribution, and accurate thermal measurements following the standard techniques used in industry for TIM performance characterization. Thermal transport across the interface between the composite and the solid surface was another component of this research. The latter has practical importance for application of graphene TIMs for thermal management of power electronics.

## II.    Experimental Procedures

For this study, we synthesized non-cured silicone oil-based composites with graphene and FLG fillers. The fillers were produced from graphite using the liquid phase exfoliation (LPE) technique [67–71] in aqueous surfactant solution in combination with liquid cascade centrifugation [72,73] for size selection. This allowed us to produce three fractions from the same stock dispersion with different lateral size and thickness distributions. The exfoliated graphene fillers were extracted from the solvent and incorporated within the base polymer to produce the required composites. The details of the exfoliation and composite preparation are outlined in the METHODS. The fillers were characterized for lateral dimension and thickness, *i.e.*, number of layers, using atomic force microscopy (AFM) and optical extinction spectroscopy to assess the concentration/mass. In Figure 1, we present the AFM characterization data for the three size-selected sets of fillers. Figure 1 (a-

c) shows the two-dimensional images of the dispersed graphene fillers in the size-selected fractions. The procedure used to assess the average dimensions is outlined in the METHODS. The resultant histograms for the longest dimension of the fillers (termed length, $L$) are given in Figure 1 (d-f) while the histograms for the number of atomic planes, $N$, are given in Figure 1 (g-i). Note that due to the lognormal shape of the distributions, there are different ways to describe the average. Here, we focus on the arithmetic mean of the filler dimensions which were 1200 nm, 800 nm and 400 nm in the case of the longest dimension, while the average number of atomic planes were determined to be 40, 19 and 8. For simplicity, we label these fillers as "large", "medium" and "small," respectively. Additional characterization data for the graphene fillers, *i.e.*, scanning electron microscopy (SEM) and optical extinction spectroscopy are provided in the Supplementary Information.

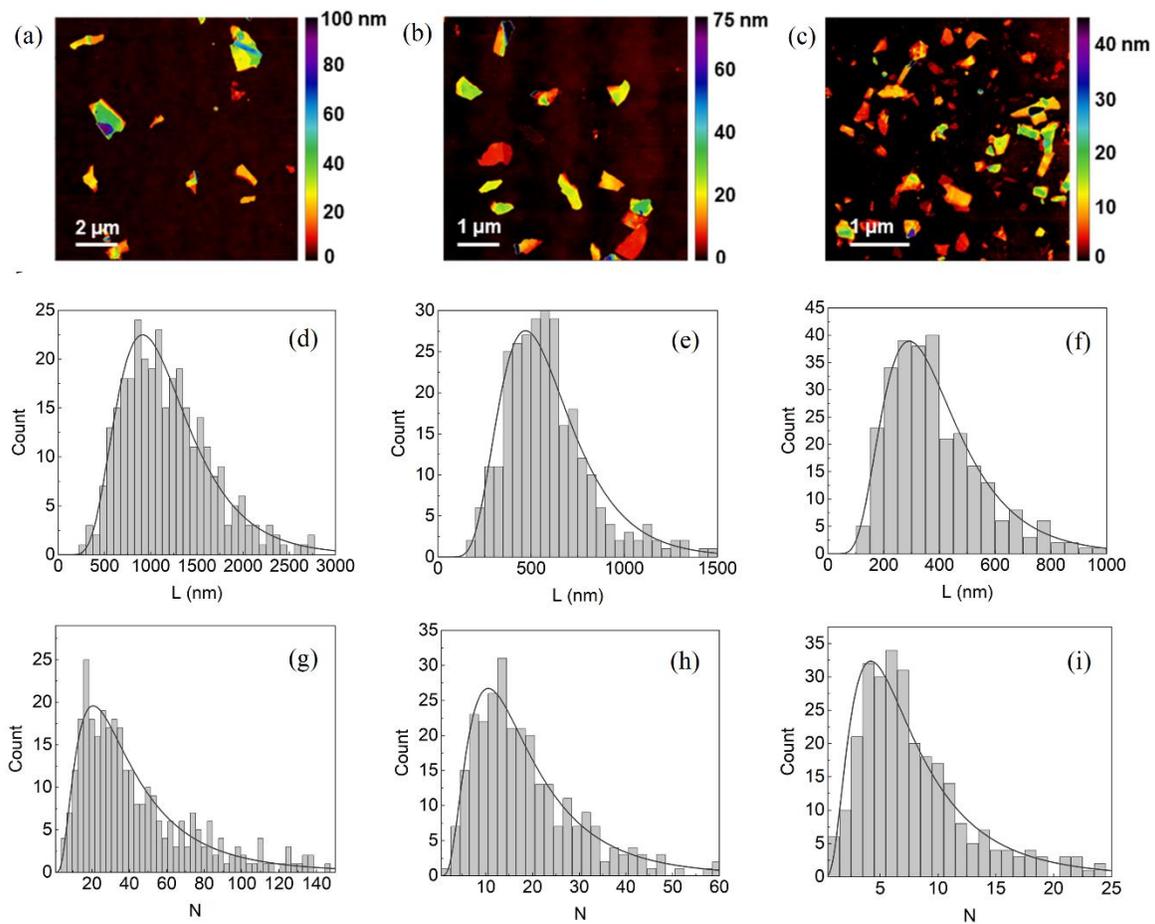

[**Figure 1:** Characterization of lateral dimensions of graphene fillers. (a–c) AFM images of the large (1200 nm), medium (800 nm) and small (400 nm) size fillers, respectively after drop-casting the dispersions on Si/SiO$_2$. (d-f) Histograms of the measured filler lengths used for determining three groups of the filler sizes. (g-i) Histograms of the number of atomic planes in the three groups of the fillers. Note that the large fillers have lateral dimensions near or exceeding the gray phonon MFP in graphene.]

It is important to note that there is a correlation between the lateral dimensions and the number of atomic planes in FLG. One cannot obtain FLG fillers with a fixed number of atomic planes and different lateral dimensions. This is related to details of the exfoliation mechanism which can be understood in terms of delamination accompanied with tearing [74]. The centrifugation, which is required for the size selection, enhances this correlation and typically produces small fillers that are thinner and large fillers that are thicker [72–74]. This correlation can be illustrated *via* a plot of the filler area as function of number of the atomic planes. Figure 2 (a) shows the measure of the characteristic lateral length, $L_* = (L \times W)^{0.5}$ *vs.* the number of atomic planes, N (here $L$ and $W$ are the measured "length" and "width" of the fillers from AFM). In this plot, each data point corresponds to a single graphene sheet measured with AFM in the size-selected fractions. The exact specification of the fillers obtained from AFM data are summarized in Table I. For FLG, the lateral dimensions affect the thermal transport stronger than the thickness [32–34,40,75–79]. For this reason, the synthesized three sets of composite samples are meaningful for understanding the specifics heat conduction and the effects of the filler size.

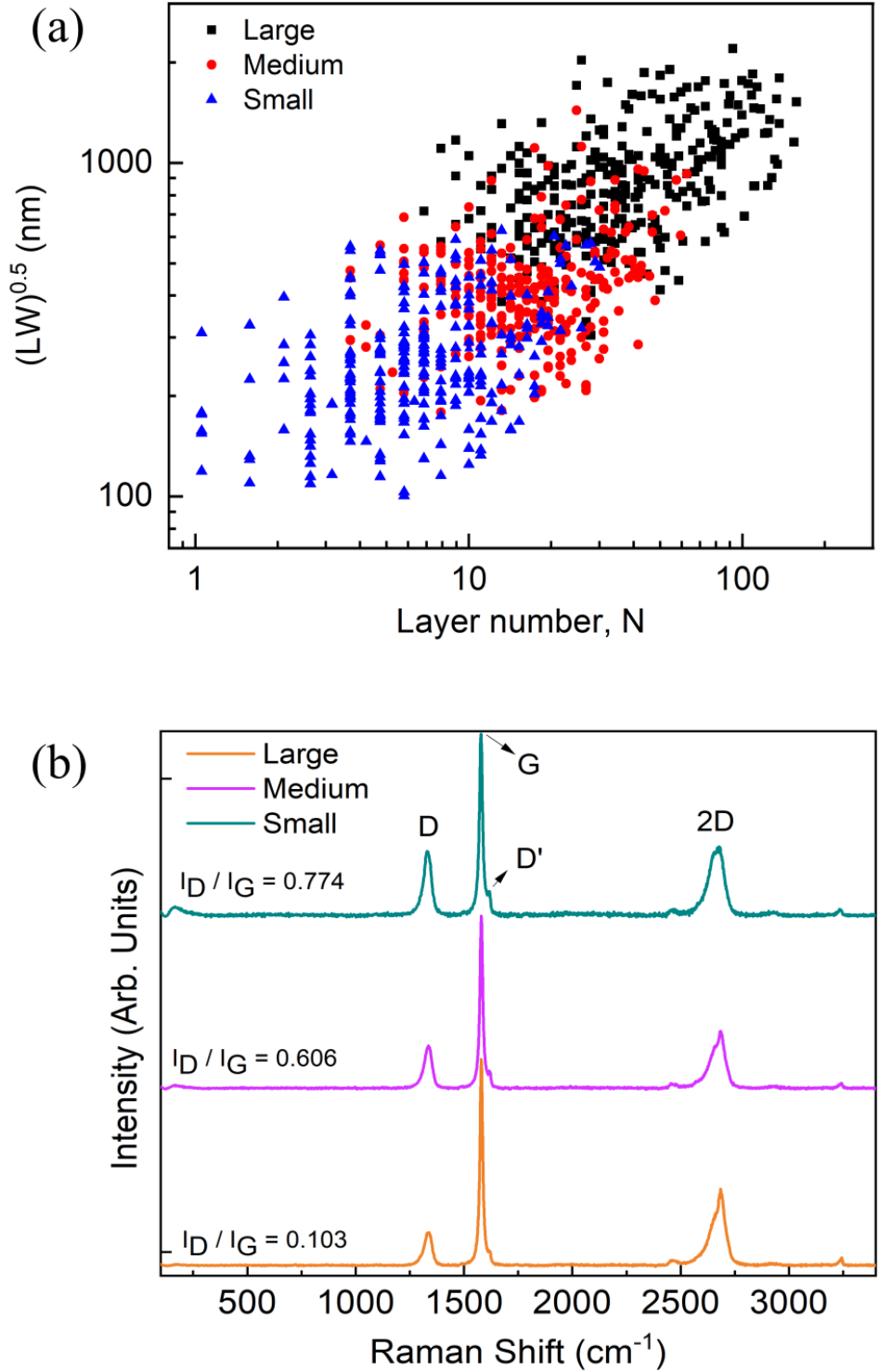

[**Figure 2:** (a) Correlation between the measure of the lateral dimensions of the fillers and the number of the atomic planes. The numerous data points denote the multiple measurements conducted with AFM. (b) Raman spectra of the graphene fillers of different lateral dimensions.]

We have verified the quality of fillers using Raman spectroscopy (Renishaw inVia). Figure 2 (b) shows the measured Raman spectra of the size-selected graphene samples used in this study. To conduct the tests, a small portion of the fillers were transferred onto a Si/SiO$_2$ substrate. The light scattering spectra were collected under a laser excitation of the wavelength 633 nm (red) and an excitation power of 2 mW at room temperature. The Raman spectra show well-known signatures of FLG, which are the G peak and 2D band [73,75,76,80–82]. As expected, the intensity of the disorder D peak increases as the size of the fillers decrease. These evolution of the peaks with the lateral dimensions and the number of the atomic planes is in line with prior literature reports [80,81,83–85]. The I$_D$/I$_G$ ratio of the spectra can be qualitatively used to describe the difference in the average size of the fillers [86]. The I$_D$/I$_G$ ratio decreases as the size of the fillers increases, changing from 0.774 for the small fillers to 0.103 for the large fillers. The evolution of the D peak and I$_D$/I$_G$ can be explained by the relaxation of the selection rules, which prohibit the appearance of the D peak in graphene with perfect translation symmetry, without defects or edges [87,88]. In the samples with smaller graphene fillers, the excitation laser light covers more graphene fillers with the edges that act as inherent defects thus resulting in an increased I$_D$/I$_G$ ratio. The 2D peak becomes more symmetric with decreasing dimensions of the fillers which is in this case related to change in the number of atomic planes. However, as outlined recently, this cannot be used as quantitative measure for thickness in LPE graphene [89].

[**Table I:** Characteristics of Graphene – FLG Fillers. Arithmetic averages of the dimensions were calculated from the AFM size statistics.]

| | Symbol | Large | Medium | Small |
|---|---|---|---|---|
| Average length (nm) | $L$ | 1200 | 600 | 400 |
| Aspect ratio (length/width) | $L/W$ | 1.8 | 2.1 | 2.3 |
| Aspect ratio (length/thickness) | $(L/t)$ | 110 | 120 | 200 |
| Average layer number | $N$ | 40 | 19 | 8 |
| Characteristic length $(L \times W)^{0.5}$ (nm) | $L_*$ | 900 | 430 | 270 |

The bulk thermal conductivity, total thermal resistance, $R_{tot}$, and thermal contact resistance, $R_C$, of the composites with the steel plates were measured following the ASTM D5470-06 standard with an industrial TIM tester (LongWin Science and Technology Corp.), which utilizes the steady-state method [90]. The measurement setup of the equipment is comprised of two polished flat steel plates, with the nm-scale roughness, as the heat source and the sink [28]. The composites were applied and tested between these plates while the heat flow and the temperature of the source and sink were carefully controlled. The thermal conductivity of the composite was extracted using the one-dimensional Fourier heat transport equation for a given $BLT$ of TIM. The experimental setup measures the total thermal resistance defined as [10,11,91]:

$$R_{tot} = \frac{BLT}{K_{TIM}} + R_{C1} + R_{C2} \tag{1}$$

Where, $K_{TIM}$ is the thermal conductivity of the composite and $R_{C1}$ and $R_{C2}$ are the thermal contact resistances of the composite with the two steel contact surfaces. The measurements were performed on composites with different loading of graphene for all the three filler sizes. The thickness of the composites was controlled using the plastic shims. It is important to note that the thermally insulating shims occupy a negligible portion of the area and volume of the composite and their contribution to overall heat transfer is negligible. All measurements have been performed under an applied pressure $P = 0.55$ MPa (~80 psi). Further details of the thermal testing are provided in the METHODS.

## III.     Results and Discussion

The total thermal resistance, $R_{tot}$, of the prepared non-cured graphene composites as a function of $BLT$ is presented in Figure 3(a-c). The data are shown for all three lateral dimensions of the graphene fillers at different loading fractions, $f$. As expected, the total thermal resistance increases linearly with $BLT$. The acquired data were used to plot a linear regression fitting for each loading fraction and for each size of the fillers. The inverse of the line slope was utilized to determine the bulk thermal conductivity of the composite using the formula: $K_{TIM}$ (Wm$^{-1}$K$^{-1}$) = $(1/Slope) \times 100$, while the $y$-intercept of the fitted line provided the thermal contact resistance, $2R_C$, assuming that $R_{C1} = R_{C2} = R_C$. One can see that the thermal resistance at constant thickness is the highest for the pure matrix material, $i.e.$, silicone oil, without the graphene addition. It is the lowest for the composites with the higher loading fraction of graphene. However, it is difficult to make conclusions about the effect of the filler size from this plot. The latter can be done in a meaningful way from the analysis of the thermal conductivity and contact resistance.

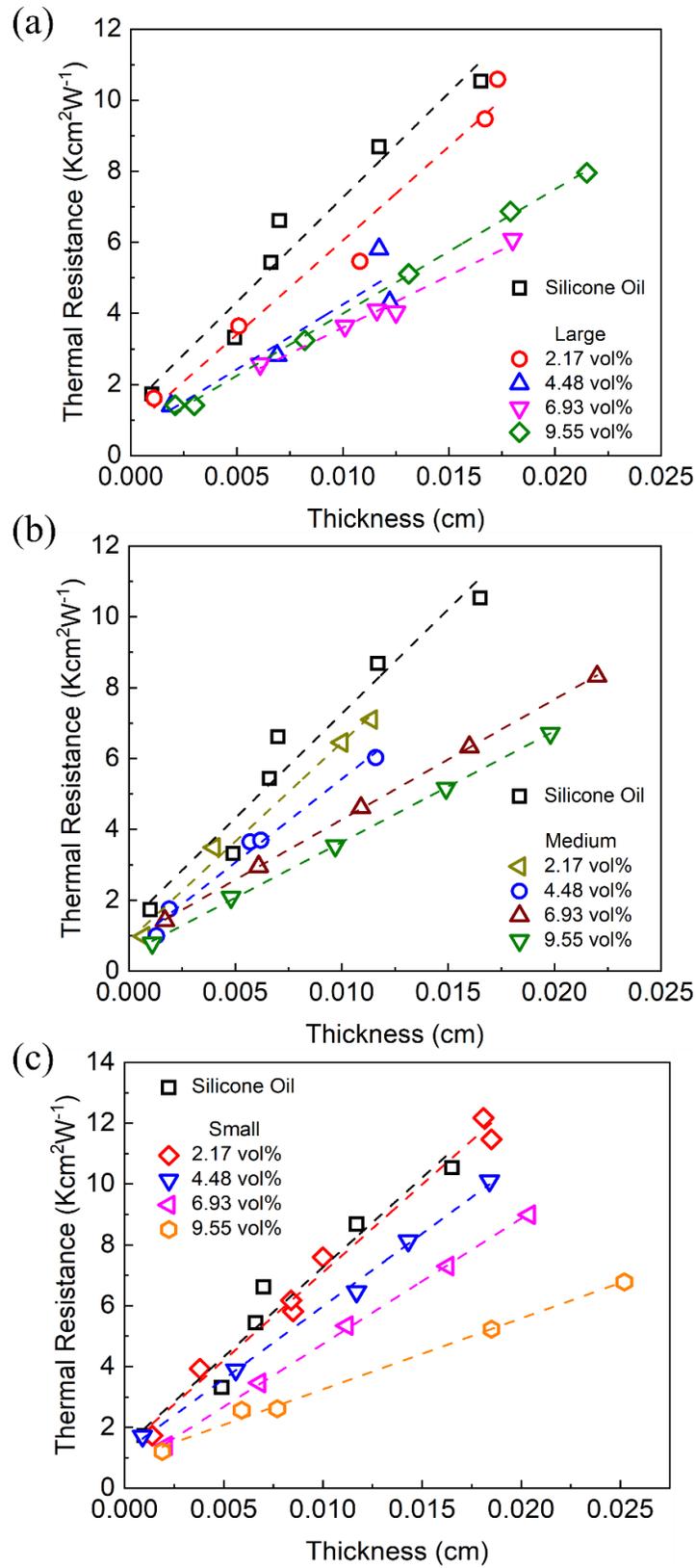

[**Figure 3:** Total thermal resistance of the graphene composites as a function of the bond line thickness of the composite, *BLT*. The thermal conductivity and contact resistance were extracted from the dashed lines which show the linear regression to the experimental data. (a) Large (b) Medium and (c) Small]

Figure 4 shows the thermal contact resistance, $R_C$, as a function of graphene loading fraction for three examined graphene filler sizes. The thermal contact resistance is the smallest for the graphene fillers with the largest lateral dimensions and decreases as fillers are added to the base polymer matrix. We can rationalize the result obtained for contact resistance of fillers with different sizes using the semi-empirical model which is represented as [92,93]:

$$R_{C_1+C_2} = 2R_C = c\left(\frac{\zeta}{k_{TIM}}\right)\left(\frac{G}{P}\right)^n. \tag{2}$$

In Eq. (2), $G = \sqrt{G'^2 + G''^2}$, where $G'$ and $G''$ are the storage and loss shear modulus of the TIM, $P$ is the applied pressure, $\zeta$ is the average roughness of the two binding surfaces and $c$ and $n$ are empirical coefficients, respectively. It can be seen that with a constant applied pressure, the prediction of thermal contact resistance becomes burdensome as the two parameters have opposing effects on the result due to the fact that addition of graphene fillers would result in an increase of both $k_{TIM}$ and $G$. This is true irrespective of TIMs with varying filler dimensions. However, this equation also suggests that for TIMs with a specific filler, graphene fillers of varying filler dimensions in this case, there would exist an optimum filler loading at which the "bulk" thermal conductivity, $k_{TIM}$, would increase significantly while the contact resistance, $R_c$, being slightly affected. This statement becomes more realistic if the total thermal resistance would be restated using Eq. (2) as:

$$R_{tot} = \left(\frac{1}{k_{TIM}}\right)\left\{BLT + c\zeta\left(\frac{G}{P}\right)^n\right\}. \tag{3}$$

Thus, with both Eq. (2) and (3), we can explain the importance of increasing the TIM bulk thermal conductivity in order to reduce the total thermal resistance. Another important aspect of non-cured TIMs that has to be considered in practical applications is the issue of pump-out, which is dependent on pressure and temperature [23,94–96]. There have been reported studies on $G'$ and $G''$ values in composites of varying filler sizes. [97–106] It has been shown that $G'$ and $G''$ decrease as the filler size increases. The viscosity of the base polymer is also an important parameter, which affects $G'$ and $G''$ and, ultimately, the pump-out effect [91,107,108]. It has been suggested [93] that $G'$ of TIM should be greater than $G''$ to avoid the issue of pump-out.

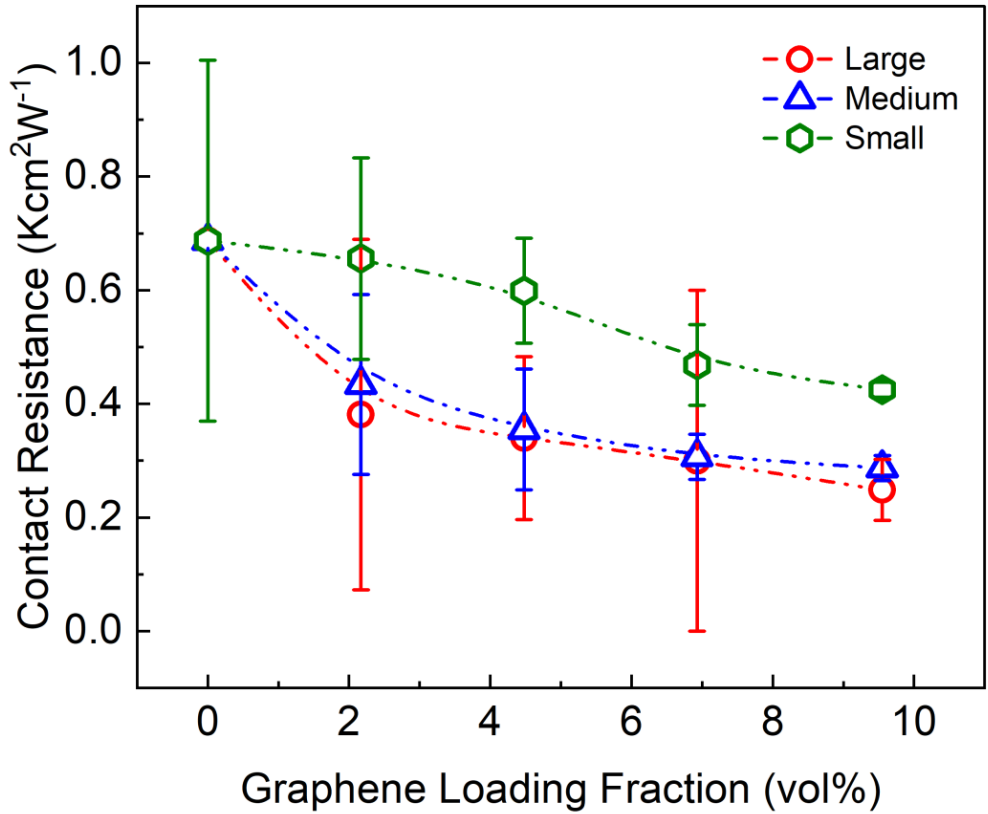

[**Figure 4:** Contact resistance of the composites as a function of the graphene loading fraction. The error bars show the standard error derived from the linear regressions. The dashed lines are shown as guides to the eye.]

In Figure 5, we present the extracted experimental thermal conductivity as a function of graphene loading fraction for three examined graphene filler sizes and its comparison with theoretical modelling. The measured thermal conductivity of the base polymer matrix – silicone oil was $0.17 \, \text{Wm}^{-1}\text{K}^{-1}$ which agrees with the previously reported values [28,109–114]. The thermal conductivity increases slightly till a loading fraction of $f = 2.17 \, \text{vol}.\%$ and then start to increase faster but still linearly. The thermal conductivity is higher for the large filler size at all the measured loading fractions. One can clearly see that in the examined range of the lateral dimensions the thermal conductivity of the composites increases with the increasing size of the graphene fillers. We rationalize this trend by comparing the size of the graphene fillers with the average, also referred to as *gray*, phonon MFP in graphene. The gray phonon MFP in graphene is in the order of ~800 nm at RT [40,75–79,115]. The enhancement of the thermal conductivity of the composites stems from thermal transport, which happens, at least partially, via the graphene fillers. If the filler lateral dimensions become smaller than the phonon MFP its own thermal conductivity decreases. The simple estimate comes from the Debye model, where the thermal conductivity $K \sim C v \Lambda$ (here $C$ is the specific heat, $v$ is the average phonon group velocity and $\Lambda$ is the average phonon MFP). When the size of the filler $L < \Lambda$, the thermal conductivity, $K$, scales down linearly with lateral dimension, $L$. This explanation remains valid for the thermal transport regime below and above the thermal percolation threshold [24].

To describe the effect of filler content on thermal conductivity of a polymer composite in Figure 5, the Kanari model [116] has been applied. This model was originally introduced as an empirical extension of the Bruggerman relation [117]. The derivation of the Kanari model within the micromechanical framework was described in details previously [118]. According to this approach, thermal conductivity, $K_{TIM}$, of a composite consisting of a polymer matrix reinforced with particles of an arbitrary shape is determined by the equation:

$$\frac{k_f - K_{TIM}}{k_f - k_m} (\frac{k_m}{K_{TIM}})^{\frac{1}{B}} = 1 - f, \tag{4}$$

Where $k_m, k_f$ are the thermal conductivities of the matrix and filler, respectively, $f$ stands for volume fraction of filler, and $B$ is a parameter characterizing shape of the filler particles. We treat these inclusions (stacks of graphene plates) as oblate ellipsoids of rotation with semi-axes $a_1 < a_2 = a_3$, where $a_1$ denotes the characteristic thickness of a stack, and $a_2 = a_3 = L_*$ stands for its in-plane size. Under this condition and given that $k_f \gg k_m$, the coefficient $B$ reads

$$B = \frac{4 - 3M}{3M(1 - M)} \tag{5}$$

with

$$M = \frac{2\varphi - \sin\varphi}{2\sin^2\varphi}\cos\varphi, \qquad \cos\varphi = \frac{a_1}{a_2} \tag{6}$$

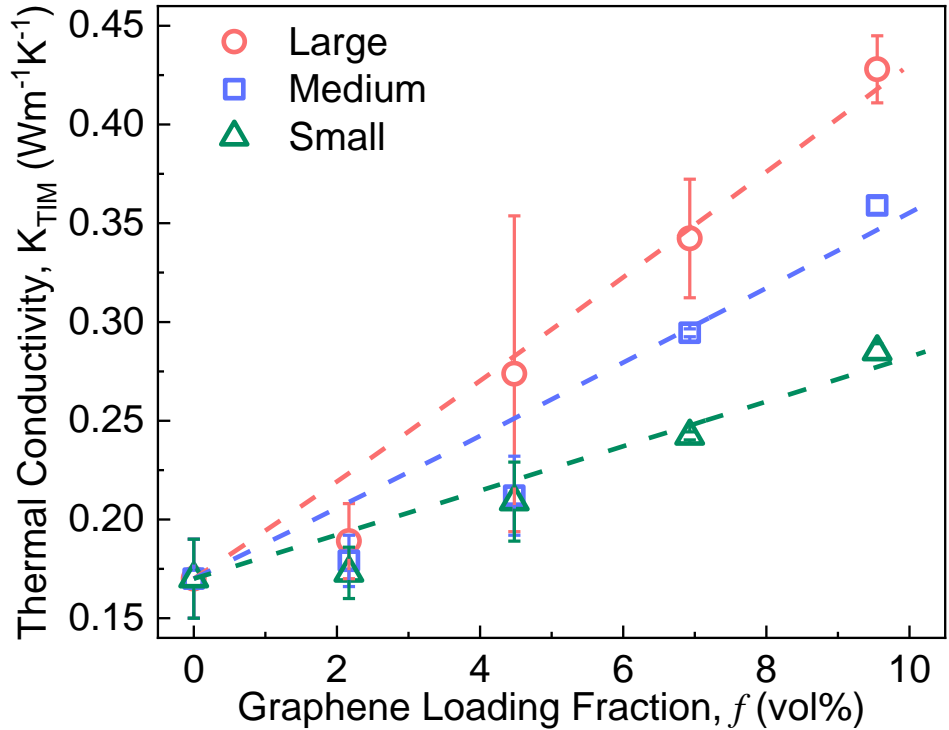

[**Figure 5:** Experimentally acquired thermal conductivity (symbols) of the non-cured composites as a function of the graphene loading fraction, $f$, for three different sizes of the fillers with comparison of Kanari model (dashed lines).]

Figure 5 demonstrates reasonable agreement between the experimental data of composites with stacks of small, medium and large graphene fillers and predictions of Eq. (4). Each set of data is matched separately by means of the only parameter: the effective thermal conductivity of stacks $k_f$. In simulation, we adopt the value $k_m = 0.17\ \mathrm{Wm^{-1}K^{-1}}$ for thermal conductivity of the matrix. The coefficients $\varphi$, $M$, and $B$, related to the aspect ratio of stacks, are calculated based on the experimental characterization of the fillers (Table I) and their values are summarized in Table II. The experimentally obtained set of data for TIMs with large, medium and small size graphene fillers have been presented in the supplementary document.

[**Table II:** Simulation parameters used in Kanari model]

| Filler Type | Layer Number ($N$) | thickness ($a_1$)[nm] | Plane dimension ($a_2 = a_3 = L_*$)[nm] | $\varphi$ [rad] | M | B |
|---|---|---|---|---|---|---|
| | | | | | | |
| Large | 40 | 14 | 900 | 1.5552 | 0.0164 | 81.5419 |
| Medium | 19 | 6.65 | 430 | 1.55533 | 0.0163 | 82.0100 |
| Small | 8 | 2.8 | 270 | 1.56043 | 0.0110 | 121.5660 |

The effect of the in-plane size of fillers on their effective thermal conductivity $k_f$ is illustrated in Figure 6. In this figure, the $k_f$ is plotted versus $L_*$ together with approximation of the data by an exponential function

$$k_f = k_f^0 + k_f^1 \exp(-\alpha L_*), \tag{7}$$

where $k_f^0 = 3.32$, $k_f^1 = -4.83$, and $\alpha = 0.0033$ are fitting parameters. Figure 6 shows that $k_f$ increases by a factor of 2.2 with $L_*$ when the in-plane size of the fillers grows from 270 to 900 nm. The extracted effective thermal conductivity of fillers is orders of magnitude smaller than the intrinsic thermal conductivity of graphene due to the filler-polymer thermal boundary resistance. However, the decrease in the effective thermal conductivity of the fillers as their size shrinks below graphene's phonon MFP is originated from a strong suppression of the its intrinsic thermal conductivity due to the enhanced phonon-edge scattering.

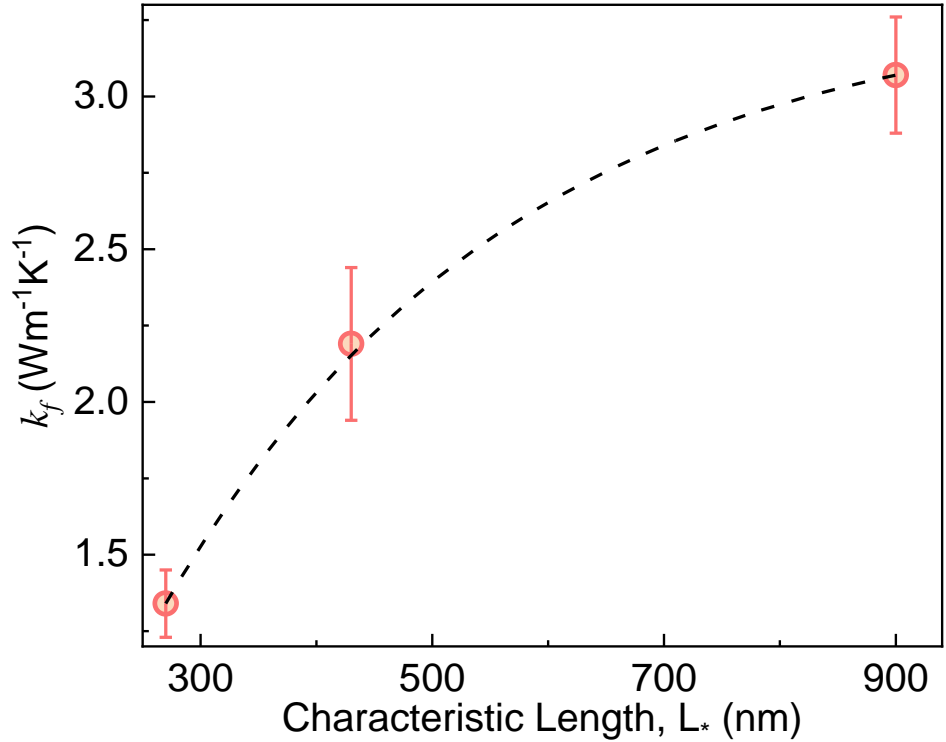

[**Figure 6:** Effective thermal conductivity of filler $k_f$ versus the characteristic in-plane size of graphene fillers $L_*$.]

Our present detailed systematic study proves that a stronger enhancement of the thermal conductivity of graphene composites can be achieved when the fillers with larger lateral dimensions are used. This conclusion is in line with our own early observation [20], and a few other reports that were performed with different graphene filler sizes [42,56,57,119–121]. The prior studies were focused on substantially larger fillers, with dimensions typically above 5 μm. In many cases, no accurate assessment and averaging were performed for the size distribution of the fillers. By selecting the average filler sizes near the phonon MFP in graphene, *i.e.,* slightly smaller, almost equal to MFP, and slightly larger, we were able to connect the observed trend with the intrinsic heat conduction properties of FLG. The observation that the thermal contact resistance also decreases with increasing the graphene filler size is non-obvious, and to some degree, it is counter-intuitive. The obtained trends shed light on the specifics of heat conduction in graphene composites, and can be used for optimization of graphene TIMs for practical applications.

## IV.     Conclusions

We reported on thermal properties of non-cured silicone composites with graphene fillers of different lateral dimensions. Graphene fillers are comprised of FLG flakes with lateral sizes in the range from 400 nm to 1200 nm and number of atomic planes from one to ~100. It was found that in the examined range of the lateral dimensions the thermal conductivity of the composites increases with the increasing size of the graphene fillers. The thermal contact resistance of composites with graphene fillers of 1200-nm lateral dimensions was also smaller than that of composites with graphene fillers of 400-nm lateral dimensions. The effects of the filler loading fraction and the filler size on the thermal conductivity of the composites were rationalized within the Kanari model. The obtained results are important for optimization of graphene fillers for applications in TIMs for heat removal from high-power-density electronics where the thermal resistance between the device and packaging is often a bottleneck impeding the heat removal. The high-power switches implemented with diamond and other wide-band-gap semiconductors are also often characterized by higher roughness allowing for the use of TIMs with larger fillers.

## METHODS

**Liquid Phase Exfoliation:** Graphene dispersions were prepared by probe sonicating the graphite powder (Sigma Aldrich 496596) with an initial concentration 40 gL$^{-1}$ in an aqueous sodium cholate (SC) solution. The powder was immersed in 80 mL of aqueous surfactant solution ($C_{surf}$= 6 gL$^{-1}$) in a stainless-steel beaker. The mixture was subjected to a two-step sonication which serves the purpose to remove impurities present in the powder in the first step: 1) The mixture was sonicated for 1 h at 60 % Amplitude in a 8 s on, 2 s off pulse sequence, using a Sonics Vibracell VCX 500 (500 W), equipped with a threaded probe and a replaceable tip. The dispersion was kept at 5 °C in a cryostat cooled water bath to avoid heating of the sample during sonication. After the sonication, the dispersion was centrifuged at 3800 $g$ for 1.5 h in a Hettich Mikro 220R centrifuge, equipped with a 1016 fixed-angle rotor at 15 °C. The supernatant containing the water-soluble impurities was discarded and the sediment redispersed in 80 ml of 2 gL$^{-1}$ aqueous surfactant solution. 2) This dispersion was then again sonicated for 5 h at 60 % Amplitude in a 6 s on, 2 s off pulse sequence.

To select nanosheets by size, we used liquid cascade centrifugation with sequentially increasing rotation speeds (2 h each step, 15°C). After each centrifugation step, supernatant and sediment were separated, the sediment collected in reduced volume (~10 mL) and surfactant concentration (0.1 gL$^{-1}$) and the supernatant subjected to centrifugation at higher speeds. Centrifugation conditions are expressed as relative centrifugal field (RCF) in units of the earth's gravitational force, *g*. Centrifugation was performed in a JA25.50 fixed angle rotor of a Beckman Coulter Avanti centrifuge at 15°C for 2 h with 50 mL centrifuge tubes (VWR, order number 525-0402) filled with 20 mL of dispersion each. The following RCF were used: 30, 200, 2500 *g*. The supernatant after 2500 g was discarded. The sediment after the centrifugation step at 30 g contains the largest sheets, but also unexfoliated bulk material. To separate the two, the dispersion was left to settle for 16 h and the top 80% of the volume decanted and collected. This sample is denoted as "large". "Medium" refers to the sample trapped between 30 and 200 *g* centrifugation, while "small" denotes the sample obtained from centrifugation between 200 and 2500 *g*. The exfoliation and size selection procedure were repeated 4 times to produce a sufficiently large mass of nanosheets. The same fractions of all batches were combined. In a final step, the filler was transferred to t-Butoxymethyl-oxiran as a solvent for better compatibility with the composite preparation. To this end, the combined fractions were centrifuged at 3000 *g* for 2h, the clear supernatant removed and the sediment redispersed in the solvent.

**Characterization of Fillers:** Optical extinction was measured on a Varian Cary 6000i in quartz cuvettes with a pathlength of 1 cm in 0.5 nm increments and integration times of 0.1 s/nm. The obtained sediments after LCC were diluted with an aqueous surfactant solution containing (0.1 gL$^{-1}$) such that their optical densities after background subtraction were < 1. The extinction at 750 nm was multiplied with the dilution factor and the concentration determined using the reported size-independent extinction coefficient of 54.5 gL$^{-1}$cm$^{-1}$ [73]. With knowledge of the total volume, the total mass of graphene in each sample was calculated. For AFM, the concentrated dispersions after size selection were diluted with de-ionized water to optical densities of ~1 and drop-cast on Si/SiO$_2$ wafers that were heated to ~150°C on a hotplate. This results in flash evaporation of the solvent to minimize aggregation on the substrate. For imaging, a Dimension ICON3 scanning probe microscope (Bruker AXS S.A.S.) was used in ScanAsyst mode (non-contact) in air under ambient conditions using aluminium coated silicon cantilevers (OLTESPA-R3). Typical image sizes ranged from 20x20 $\mu m^2$ in the case of the largest sheets to 5x5 $\mu m^2$ for the sample containing the smaller sheets, each scanned with 1024 lines at scan rates of ~0.4 Hz and peak force setpoints of ~0.1 V. Using Gwyddion software, the longest dimension (length), dimension perpendicular (width) and thickness of ~300 individually deposited nanosheets in each sample was manually measured. The thickness was converted to layer number using the reported step height of LPE graphene of 0.9 nm [122]. The lateral dimensions were corrected for cantilever broadening and pixilation by using an empirical correction [123]. From the distributions, the arithmetic averages of the dimensions were calculated. Due to the lognormal shape of the distributions, we note that the arithmetic average is larger than the maximum (i.e. the mode) of the distribution. To estimate the average length/thickness aspect ratio, the length of each sheet was divided by the layer number multiplied with the crystallographic thickness of one layer (0.35 nm) and the arithmetic mean calculated.

**Composite Synthesis:** The exfoliated graphene fillers were centrifuged with a solvent of t-Butoxymethyl-oxiran in a vial at 7000 rpm for 2 minutes to ensure that the fillers were visually separated from the solvent. The solvent was removed from the vial to an extent using a dropper and the remaining contents of the vial were transferred to a petri dish in a fume hood for 24 hours at room temperature. This enabled the solvent to completely evaporate leaving us with only the fillers. These extracted fillers were precalculated and added to the base polymer matrix of silicone oil (Fisher Scientific, USA), also known as PDMS – Poly(dimethylsiloxane). The obtained compound was then shear mixed (Flacktek Inc.) at a speed of 2000 rpm for 10 minutes followed by a speed of 3000 rpm for 15 minutes to ensure homogenous distribution of the filler in the polymer. This process was repeated for all the filler sizes and with all loading fractions. Raman spectra of the polymer matrix and prepared composites at 6.93 vol% are provided in the Supplementary Information.

**Thermal Characterization:** The thermal conductivity and thermal contact resistance of the composites were measured using the industrial TIM tester (LongWin Science and Technology Corp, Taiwan) utilizing the steady-state method according to the requirements of industry standard ASTM D5470-06. The noncuring graphene composites were tested under the pressure of 80 psi and a temperature of 80 ℃ for 40 min for each thickness using plastic shims (Precision Brand Products Inc, USA).

### Supporting Information

Supporting Supplementary Information is available from the journal web-site for free of charge.

### Acknowledgements

The work at UC Riverside was supported, in part, by ULTRA, an Energy Frontier Research Center (EFRC) funded by the U.S. Department of Energy, Office of Science, Basic Energy Sciences under Award #  DE-SC0021230.

### Contributions

F.K. and A.A.B. conceived the idea of the study and coordinated the project. S.S. prepared the thermal compounds, performed thermal conductivity measurements, thermal contact resistance measurements and conducted data analysis of thermal and Raman data; F.R. performed LPE, AFM and optical extinction measurements on the fillers and conducted analysis; D.W. assisted with thermal measurements and analysis; Z.B. performed Raman spectroscopy measurements and assisted in thermal compound preparation and analysis of Raman data; A.D.D performed modeling of the experimental data and compared them; F.K. contributed to the experimental and theoretical data analysis; S.B. performed SEM of the graphene fillers; C.B. supervised LPE, AFM and optical absorption of the fillers and contributed to analysis; A.A.B. contributed to the thermal data analysis; F.K. and A.A.B. led the manuscript preparation. All authors contributed to writing and editing of the manuscript.